\documentclass[10pt]{article}
\usepackage[english]{babel}
\usepackage{latexsym}

\usepackage{mathtext} 		
\usepackage[utf8]{inputenc}	
\usepackage{indentfirst}
\usepackage{subfigure}
\usepackage{dsfont}

\usepackage{amsmath,amsfonts,amssymb,amsthm,mathtools} 
\usepackage{icomma}
\usepackage{tikz}
\usepackage{pgfplots}

\usepackage{graphicx} 
\usepackage{wrapfig}

\usepackage{cite}
\usepackage{csquotes}

\usepackage{array,tabularx,tabulary,booktabs}
\usepackage{longtable}
\usepackage{multirow}

\theoremstyle{plain} 
\usepackage{etoolbox} 

\usepackage{extsizes} 
\usepackage{geometry} 
\usepackage{setspace} 
\usepackage{lastpage} 
\usepackage{soulutf8} 

\usepackage{multicol}
\title{\LARGE \bf
Conformalized density- and distance-based anomaly detection in time-series data}

\author{
E.V. Burnaev$^{1}$,
V.I. Ishimtsev$^{2}$,\\
$^{1}$IITP RAS, Skoltech, e.burnaev@skoltech.ru,\\
$^{2}$IITP RAS, vladislav.ishimtsev@gmail.com
}

\date{9 July 2016}

\begin{document}

\maketitle
\thispagestyle{empty}
\pagestyle{empty}

\begin{abstract}
Anomalies (unusual patterns) in time-series data give essential, and often actionable information in critical situations. Examples can be found in such fields as healthcare, intrusion detection, finance, security and flight safety. In this paper we propose new conformalized density- and distance-based anomaly detection algorithms for a one-dimensional time-series data. The algorithms use a combination of a feature extraction method, an approach to assess a score whether a new observation differs significantly from a previously observed data, and a probabilistic interpretation of this score based on the conformal paradigm.
\\
\textbf{Keywords:} Non-conformity measure, anomaly detection, time-series, feature extraction, LOF, LoOP
\end{abstract}

\section{Introduction}
Anomaly detection in time-series data is an important task in many applied domains \cite{twitter}. For example, anomaly detection in time-series data can be used for monitoring of an aircraft cooling system \cite{intro0}, it can be applied in a health research to find unusual patterns, it can give a competitive edge to a trader. 
\\\indent Conventional anomaly detection methods identify patterns of normal behavior and declare that any data not similar to these patterns is anomalous. Time-series specifics, as well as several other factors greatly complicate the anomaly detection: \begin{itemize}
\item Usually normal behavior is described by a fixed model \cite{rw1, rw2}, which does not always reflect the reality. However, in many domains normal behavior continues to evolve and the current concept of normal behavior can not be sufficiently representative in the future;
\item Noise appears in the data, which blurs boundaries between a normal and an abnormal data and as a result, causes an increase of a prediction error;
\item When anomalies are the result of illegal actions, frauders often mask anomalous instances and they appear as normal ones;
\item The main goal of anomaly detection is often a warning, rather than detection, so it is important to detect the anomaly as soon as possible. It is clear that the standard quality metrics, such as the precision and the recall, in this case are not sufficiently informative.
\end{itemize}

\indent In this paper we consider approaches to anomaly detection in one-dimensional time-series data. Based on the abovementioned factors, we can say that even in case of a  one-dimensional time-series data the anomaly detection is a difficult task since the standard assumptions of classical change-point models may not be satisfied. E.g., a data can have long-range dependences, from which it is difficult to extract a signal \cite{rw3}; or it can contain quasi-periodic components \cite{rw4}. Thus, one has to consider the specialized methods for anomaly model selection \cite{rw5}, ensembling of anomaly detection statistics \cite{rw6}, resampling for balancing normal and abnormal classes \cite{rw7}, etc. However, all these approaches are designed to tackle  separately specific time-series peculiarities.
\\\indent Therefore, the aim of this paper is to propose a reliable non-parametric approach for anomaly detection in one-dimensional time-series data possessing a probabilistic interpretation of an anomaly score.

\section{Related work}
Most of the existing anomaly detection methods solve the abovementioned challenges only in case of domain-specific formulations of problems. E.g., these methods often rely on a time-series model and use it for prediction of future time-series values. In case of a multi-dimensional data some non-parametric methods are available, but they are primarily designed for independent observations. Let us briefly overview the main non-parametric approaches for anomaly detection in multi-dimensional data.
\\\indent Distance-based methods use a distance from a considered test point to its nearest neighbors assuming that the normal data points are close to their neighbors, while the anomalous data points are far from the normal data. For example, the sum of distances to the $k$ nearest neighbors (KNN) can be considered as an anomaly score: 
\begin{equation} \label{eq:knn_score}
anomaly\_score(x) = \sum_{o \in kNN(x)} dist(x,o),
\end{equation}
where $x$ is the new test data point.
This detector has two hyperparameters: \begin{itemize}
\item[---] $k$ is a number of considered neighbors;
\item[---] $\epsilon$ is an anomaly threshold.
\end{itemize}

\indent If the anomaly score of the test observation $x$ exceeds the anomaly threshold $\epsilon$, the test observation is declared to be anomaly. Drawbacks of this algorithm are the high sensitivity to the hyperparameter $k$ and a lack of interpretation of the anomaly score, since its value has no upper bound ($anomaly\_score(x)\in \mathds{R}_+$). Some modifications of this algorithm are discussed in \cite{knn1,knn2,knn3}.
\\\indent It is obvious that the distance-based methods perform poorly when structure of observations contains clusters of different densities. 
\\\indent Density-based methods solve the anomaly detection problem by introducing the concept of a data density. The larger the distance from the considered observation to its neighbors, the less its density is. Assumptions about the anomalous data are as follows: a normal observation density is close to the density of its nearest neighbors, while the density of an anomalous observation is significantly different from the density of its neighbors.
\\\indent Local Outlier Factor (LOF) method \cite{lof} uses inverted average distance to the $k$ nearest neighbors as a density measure:
\[
    loc\_dens_{k} (x) 
    = \left(\frac1k \sum_{o \in kNN(x)} reach\_dist_k(x,o) \right)^{-1},
\]
where
\[
    reach\_dist_{k}(x, o) 
   = \max \{ dist(x,o),\; dist(x, NN_{k}(x)) \},
\]
$NN_k(x)$ is the $k$-th nearest neighbor of $x$. Such definition of $reach\_dist(x,o)$ allows one to reduce statistical fluctuations when $x$ and $o$ are close to each other.
\\\indent Density of the considered observation is compared with the average density of its neighbors, and then the anomaly score, called $Local\;Outlier\;Factor$, is calculated:
\begin{equation} \label{eq:lof_score}
LOF_{k}(x) = \frac1{k} \sum_{o \in kNN(x)} \frac{loc\_dens_{k}(o)}{loc\_dens_{k}(x)}.
\end{equation}
If $LOF \approx 1$ we consider the observation $x$ to be normal, if $LOF \gg 1$ we consider $x$ to be anomalous.
\\\indent LOF method has the same set of hyperparameters --- $k$ and $\epsilon$, and, unfortunately, has the same drawbacks: high sensitivity w.r.t. the hyperparameter $k$ and a lack of interpretability of the anomaly score. Some modifications of this algorithm are discussed in \cite{lof1,lof2}.
\\\indent Also LOF method has a modification, described in \cite{loop}, called Local Outlier Probabilities (LoOP). This method allows to reduce the sensitivity w.r.t. the hyperparameter $k$ thanks to more strict assumptions: \begin{itemize} 
\item normal observation is centered w.r.t. its neighbors;
\item the distances from the observation to its neighbors are distributed normally (considering the positive half of the distribution). 
\end{itemize}

\indent In fact, LoOP method in its own way defines the local density of observations. Anomaly score for a new observation $x$ is limited, i.e. $anomaly\_score \in [0,1]$:
the closer the value to $1$, the more confident we are in our decision that $x$ is an anomaly.
\\\indent The main advantage of density- and distance-based methods is that they contain a few hyperparameters. However, detection results are significantly sensitive to their values. The challenge of this paper is to build reliable non-parametric anomaly detection methods based on the KNN and LOF ideas and adapt them to a time-series data.

\section{Feature Extraction}
Performance of the anomaly detection based on the density- and distance-based methods depends on the efficiency of the considered features. It is clear that when we have a one-dimensional time-series, the direct application of the considered methods to the initial data implies a strong deterioration in the  detection quality since information about a time dependence between observations is not taken into account. It is therefore necessary to consider some pre-processing of the data in order to provide a mapping of the time-series values to a multi-dimensional feature space.
\\\indent A method, proposed in the framework of the Singular Spectrum Analysis (SSA) \cite{ssa}, also known as ``Caterpillar'', provides an effective representation of a time-series data by a set of multi-dimensional vectors and allows keeping the dependent structure of the one-dimensional time-series. 
\\\indent The idea of the ``Caterpillar'' method can be described as follows. We denote by $X = (x_1,\dots,\;x_n)$ a time-series realization, by $L$, $1 <L < \frac{n}2$ a window length, and consider a matrix:
\begin{equation} \label{eq:ssa}
	\textbf{X} = \left[\begin{array}{cccc} 
				x_1 & x_2 & \cdots & x_M \\				
                x_2 & x_3 & \cdots & x_{M+1} \\
                \vdots & \vdots & \ddots & \vdots \\
                x_L & x_{L+1} & \cdots & x_{M+L-1} 
            \end{array}\right].
\end{equation}
\indent So we use the $L\times M$ matrix $\textbf{X}$, corresponding to the moment of time $M+L-1$, to characterize $M$ recent values of the time-series $X$. These values are $x_L, x_{L+1},\dots, \;x_{M+L-1}$. If a new observation $x_{M+L}$ arrives we switch to another matrix of size $L\times M$:
\[
    \left[\begin{array}{cccc} 
				x_1 & x_2 & \cdots & x_M \\				
                x_2 & x_3 & \cdots & x_{M+1} \\
                \vdots & \vdots & \ddots & \vdots \\
                x_L & x_{L+1} & \cdots & x_{M+L-1} 
            \end{array}\right] 
             \rightarrow \left[\begin{array}{cccc} 
				x_2 & x_3 & \cdots & x_{M+1} \\				
                x_3 & x_4 & \cdots & x_{M+2} \\
                \vdots & \vdots & \ddots & \vdots \\
                x_{L+1} & x_{L+2} & \cdots & x_{M+L} 
            \end{array}\right].
\]

\section{Proposed Approach}
In order to detect anomalies, we have to measure a mutual dissimilarity of observations. To solve this task, let us consider the approach, proposed  \cite{cad} and based on conformal prediction (CP) \cite{cp}. 
\\\indent The basic idea of CP can be described as follows. Using a training set $(x_1,\dots,\;x_m)$ for a new observation $x_{m+1}$ we compute the parameter $p$: probability that in the training set we can find an observation with a more extreme value of a Non-Conformity Measure (NCM) then value of the NCM for this new observation. Usually construction of an NCM is a domain-specific procedure.
\\\indent CP method is easily adjusted to the anomaly detection problem (Conformal Anomaly Detection --- CAD). It is enough to interpret parameter $p$ as a parameter of data ``normality''. 
\\\indent A high computational complexity is one of the main drawbacks of CAD. In \cite{icad} there is a modification of CAD, called Inductive Conformal Anomaly Detection (ICAD). The main idea of this method is the following: we have two data sets –-- ``proper training'' and ``calibration'' sets. For each data point from the calibration set we obtain its value of NCM using the proper training set. Also we calculate value of NCM for each data point from the test set. The parameter $p$ is then estimated by comparing these sets of NCM values, see details below.
\\\indent Let us describe the proposed algorithm.
\\\textbf{Input:} \begin{itemize}
\item window length $L$;
\item  size of the proper training set $T$;
\item size of the calibration set $C$;
\item  time-series realization $(x_1,\ldots, x_{T+C+L-1})$;
\item test observation $x_{T+C+L}$;  
\item Non-Conformity Measure $NCM$.  
\end{itemize}
\textbf{Output:} Anomaly score $p \in\;[0,1]$. 
\\\textbf{Steps:} \begin{enumerate}
\item[1.] Map the time-series realization $(x_1,\dots,\;x_{T+C+L-1})$ into the matrix $\textbf{X}\in \mathds{R}^{L\;\times\;(T+C)}$ by (\ref{eq:ssa}).
\item[2.] Split $\textbf{X}$ into the proper training matrix $\text{\textbf{X}}_T\;\in \mathds{R}^{L\times T}$ and the calibration matrix $\text{\textbf{X}}_C\;\in \mathds{R}^{L\times C}$.
\item[3.] Calculate NCM values $(\alpha_1,\dots,\alpha_C)$ for vectors $\left(\text{\textbf{X}}_C^i\right)_{i=1}^C \in \mathds{R}^L$ from the calibration matrix $\text{\textbf{X}}_C$ using the proper training set $\text{\textbf{X}}_T$:
$$
	\alpha_i = NCM(\text{\textbf{X}}_C^i, \text{\textbf{X}}_T),\;\;i=1,\dots,\;C.
$$
\item[4.] Calculate NCM value for the test observation $x_{T+C+L}$, embedded into the feature space $\mathds{R}^L$ using the proper training set $\text{\textbf{X}}_T$:
$$
	\alpha_{T+C+L} = NCM(z, \text{\textbf{X}}_T),\;\;z = \left(\begin{array}{c}
x_{T+C+1}\\
\vdots\\
x_{T+C+L}
\end{array}\right).
$$
\item[5.] Calculate the anomaly score $p$:
$$
	p = \frac{\left|i=1,\ldots, C:\; \alpha_i \geq \alpha_{T+C+L} \right|}{C}.
$$
\end{enumerate}

\indent We obtain KNN-ICAD anomaly detection method, if we use statistic (\ref{eq:knn_score}) from the distance-based KNN method as an NCM, and we obtain LOF-ICAD anomaly detection method, if we use statistic (\ref{eq:lof_score}) from the density-based LOF method as an NCM.
\\\indent Note that in the both cases we use Mahalanobis distance as the distance function $dist(\cdot,\cdot)$ in the feature space to account for mutual correlations of features.

\section{Experiments}
\label{experiments}
\begin{figure}[t!]
\centering
\includegraphics[width=\linewidth]{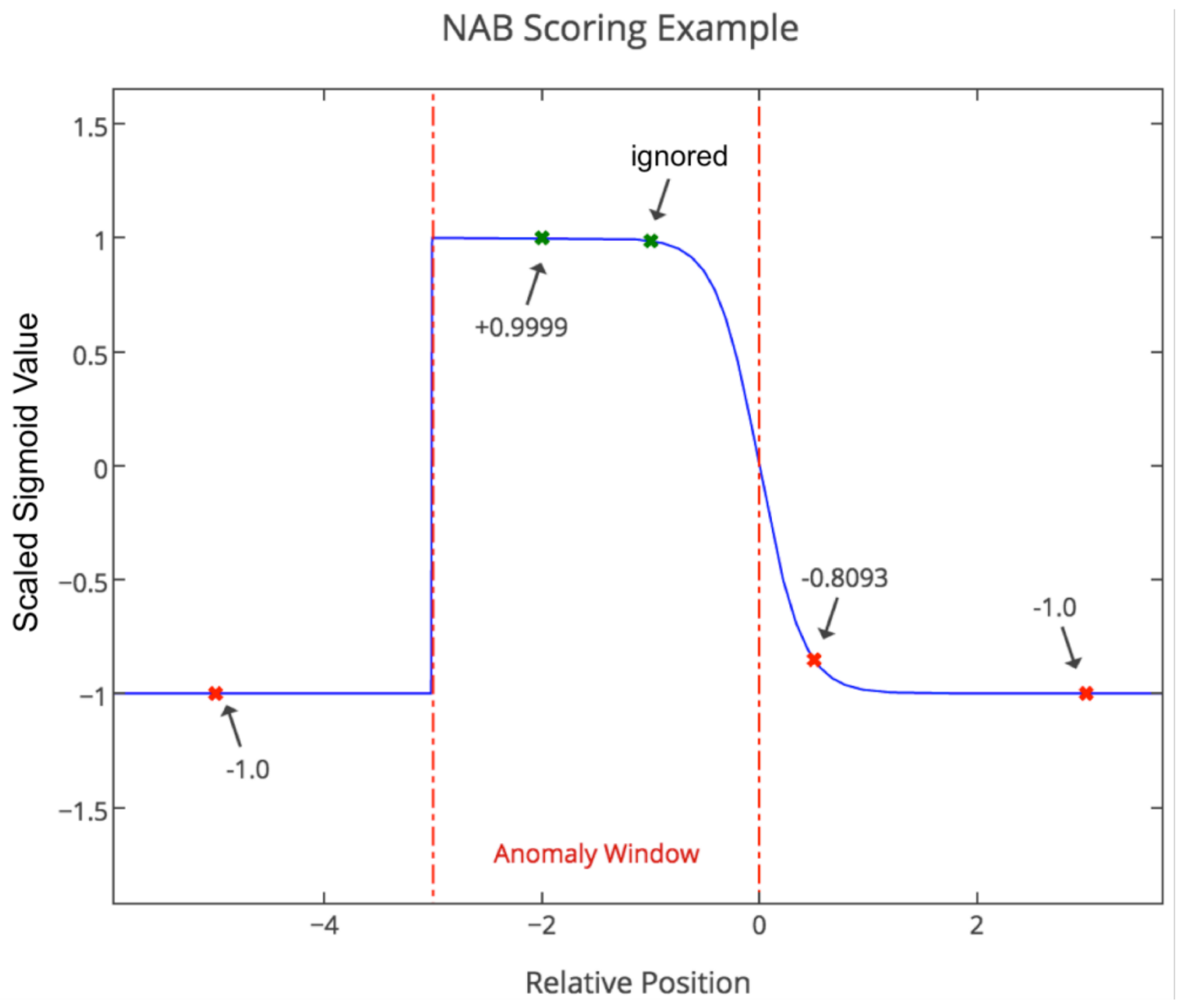}
\caption{Scoring example for a sample anomaly window, where the values represent the scaled sigmoid function. The first point is an FP preceding the anomaly window (red dashed lines) and contributes $-1.0$ to the score. Within the window we see two detections, and only count the earliest TP for the score. There are two FPs after the window. The first is less detrimental because it is close to the window, and the second yields $-1.0$ because it’s too far after the window to be associated with the true anomaly. TNs make no score contributions. The scaled sigmoid values are multiplied by the relevant application profile weight, the NAB score for this example would calculate as: $-1.0 A_{FP} + 0.9999 A_{TP} - 0.8093 A_{FP} - 1.0 A_{FP}$. With the standard application profile this would result in a total score of 0.6909 \cite{nab}.}
\label{fig:score}
\end{figure}

To test the described anomaly detection algorithms, we should use time-series, containing different kinds of anomalies. In \cite{nab} authors provide a publicly available set of 58 labeled one-dimensional time-series from different fields, called Numenta Anomaly Benchmark (NAB). NAB attempts to provide a controlled and repeatable environment of tools to test and measure different anomaly detection algorithms on streaming data \cite{nab}. 
\\\indent A measure of anomaly detection performance, proposed in NAB, takes into account the detector's responsiveness to the appearance of anomalies, and allows to set weights $A_{TP},\;A_{TN},\;A_{FP},\;A_{FN}$ for true positives, true negatives, false positives and false negatives respectively. Penalizing for missing anomalies and rewarding for the detection of anomalies is schematically shown in Figure \ref{fig:score}.\\
\begin{figure}[t!]
      \centering
      \begin{minipage}[!h]{\linewidth}
      \center{\includegraphics[width=\linewidth]{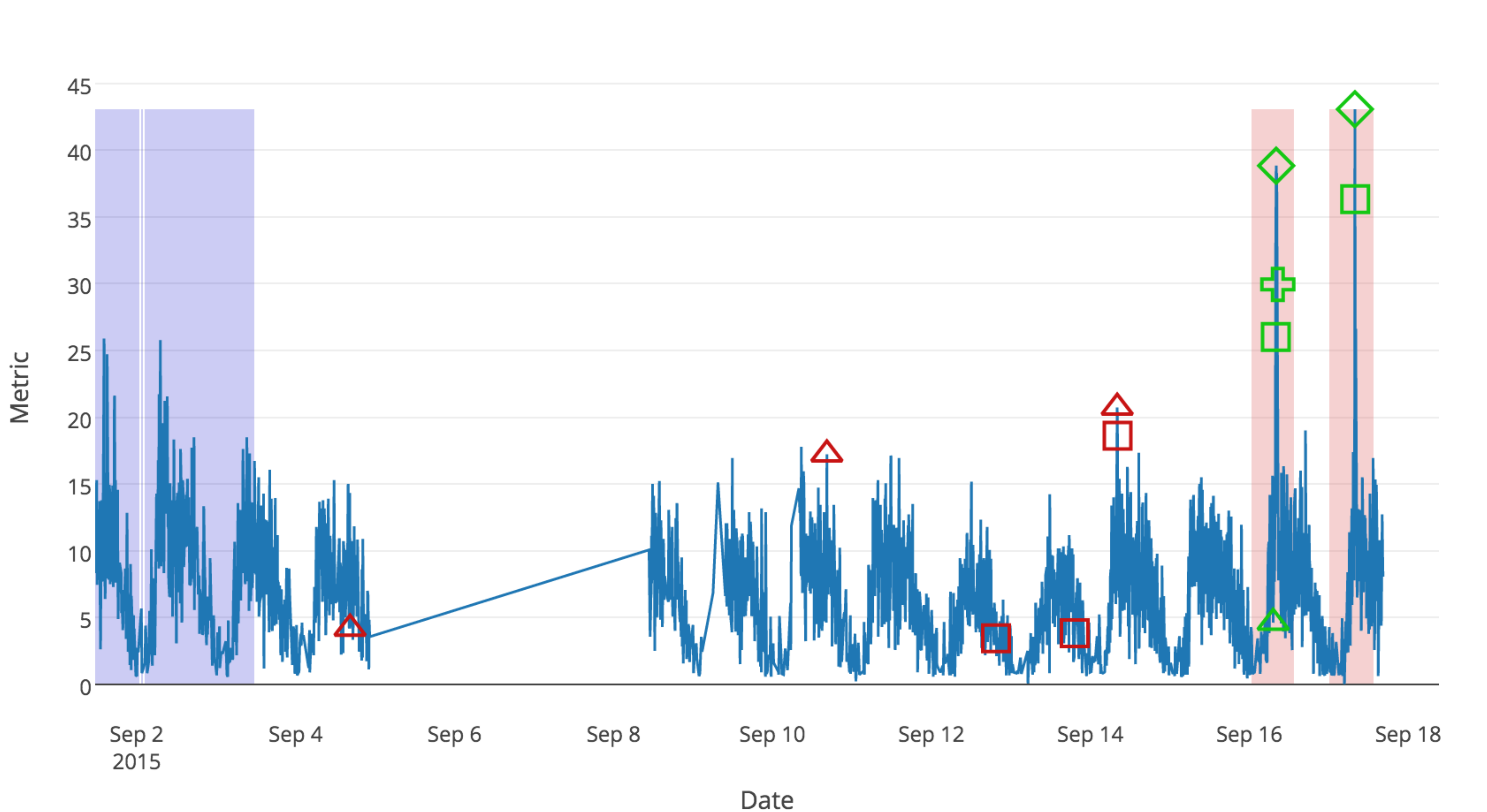}\\ }
      \end{minipage} 
      \hfill
      \begin{minipage}[!h]{\linewidth}
          \center{\includegraphics[width=0.97\linewidth]{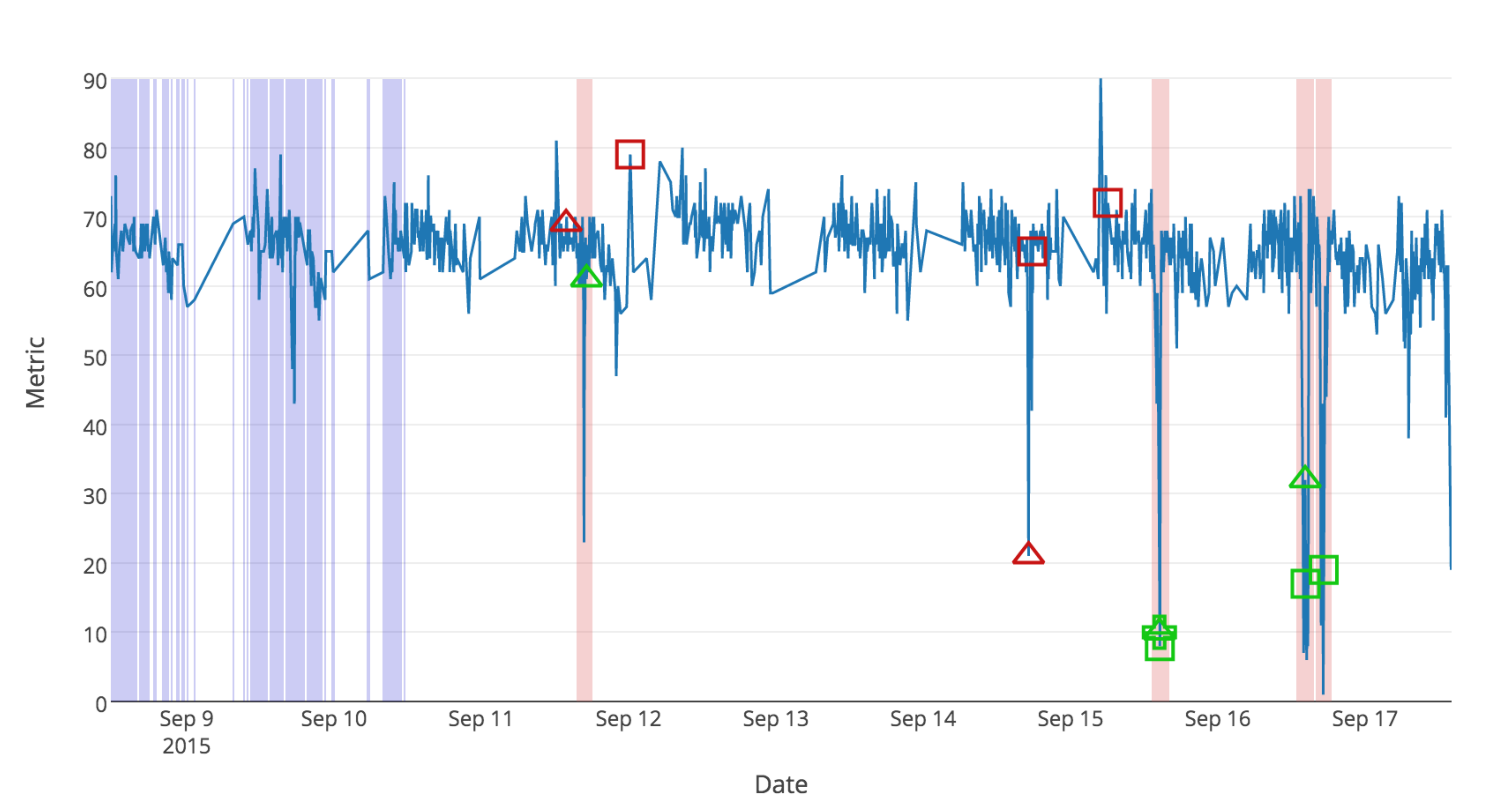}\\ }
      \end{minipage} 
      \vfill
      \begin{minipage}[!h]{\linewidth}
      \begin{center}
      \includegraphics[width=0.85\linewidth]{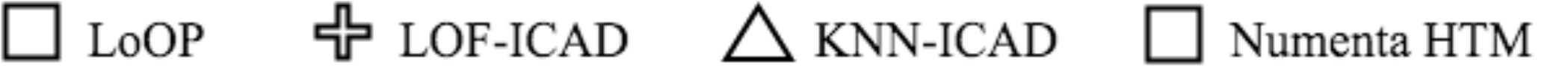}
	  \end{center}
      \end{minipage}
        \caption{Real time traffic data from the Twin Cities Metro area in Minnesota. The shaded blue region is the first 15\% of the data file, representing the probationary period. During this period the detector is allowed to learn the data patterns without being tested. For a given detector, the scored (i.e. the first) TP detection within each window is labeled in green. All FP detections are colored red. Also the red shaded regions denote the anomaly windows.} 
\label{traffic}
    \end{figure}
\indent By varying these weights, we can get different quality measures. Authors of \cite{nab} proposed to use weights, given in Table \ref{tabb1}.
\begin{table}[!h]
\centering
\caption{Weights for measuring anomaly detection performance}
\begin{tabular}{l|cccc}
Metric             & $A_{TP}$ & $A_{FP}$ & $A_{TN}$ & $A_{FN}$ \\
\hline
Standard           & 1.0      & -0.11    & 1.0      & -1.0     \\
Reward low FP rate & 1.0      & -0.22    & 1.0      & -1.0     \\
Reward low FN rate & 1.0      & -0.11    & 1.0      & -2.0    
\end{tabular}
\label{tabb1}
\end{table} \\\indent
The theoretical range of the NAB score is $final\_score \in [-\infty; 100]$, but in practice there is a lower bound depending on the number of observations in all time-series.
	\begin{figure}[t!]
	      \centering
      \begin{minipage}[!h]{\linewidth}
      \center{\includegraphics[width=\linewidth]{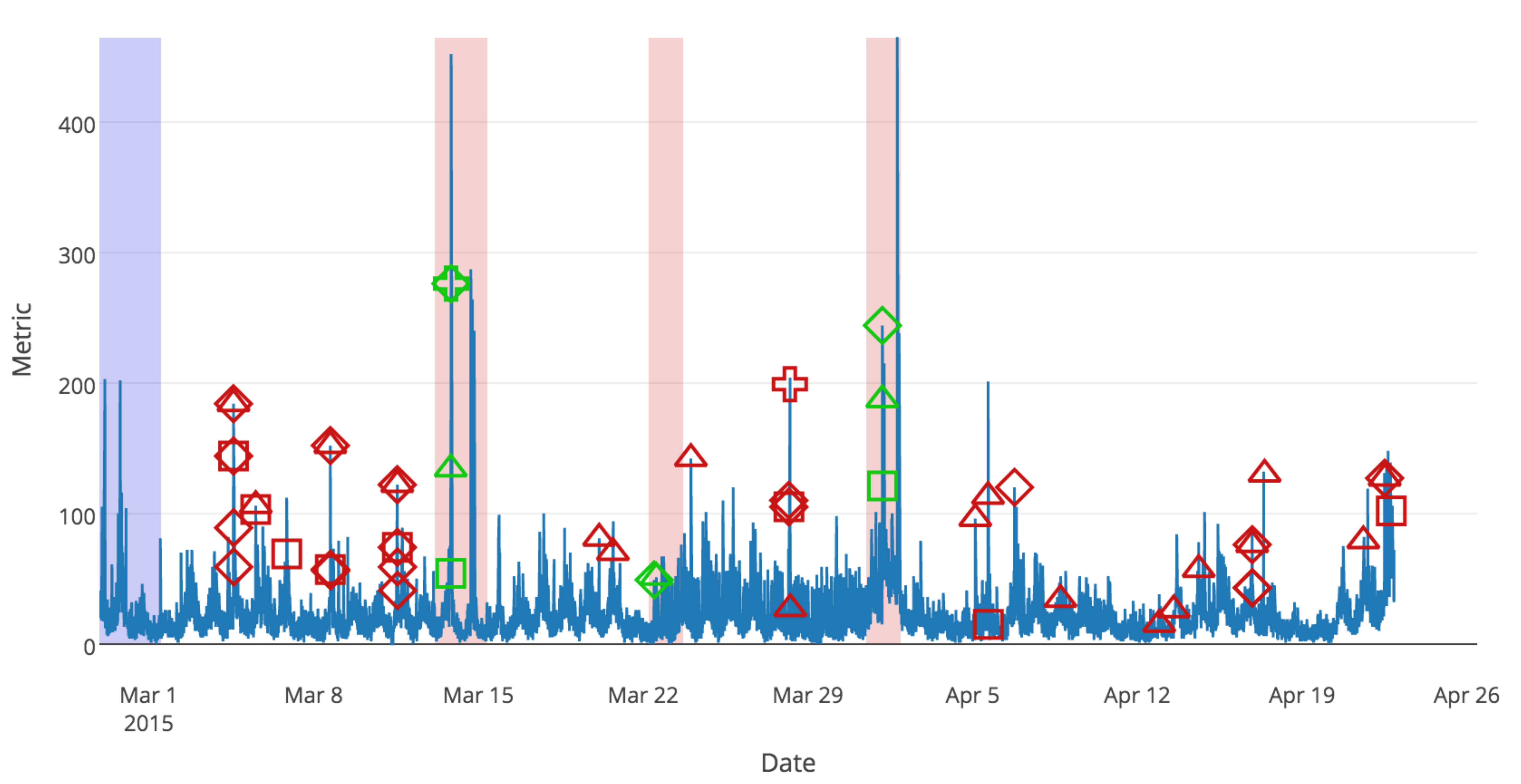}\\ }
            \end{minipage} 
              \vfill
      \begin{minipage}[!h]{\linewidth}
      \begin{center}
      \includegraphics[width=0.85\linewidth]{legend}
	  \end{center}
      \end{minipage}
        \caption{A collection of Google's Twitter mentions. The metric value represents the number of mentions for a given ticker symbol every 5 minutes. Notation of amonalies and anomaly detections are the same.} 
                \label{twitter}
	\end{figure}

\indent We should note that the Numenta Anomaly Benchmark measures performance of an anomaly detection method on the entire set of time-series assuming that the same hyperparameters are used for all test  cases.
\\\indent We compare our approach with Numenta HTM method \cite{numenta}, based on a Hierarchical Temporal Memory \cite{htm}, Twitter Anomaly Detection method \cite{twitter} and ``Null'' detector, which assumes that absolutely all data is normal.
\\\indent Twitter Anomaly Detection approach combines statistical techniques to detect anomalies. The Generalized ESD test \cite{advec} is combined with robust statistical metrics, and piecewise approximation is used to detect long term trends \cite{nab}. Note that Twitter ADVec is a domain-specific anomaly detection method (10 of 58 time-series in NAB are from Twitter).
\\\indent Examples of anomaly detections are shown in Figures \ref{traffic} and \ref{twitter}. For some time-series (Figure \ref{twitter}) conformal anomaly detection methods provide a lot of false alarm events that eventually led to decrease of the final performance. 

\indent After optimization of hyperparameters of the considered anomaly detection methods, we obtain Table \ref{tab:results}. From Table \ref{tab:results} we can see that 
\begin{enumerate}
\item The LoOP and LOF methods provides the worst results. One of the reasons is their high sensitivity w.r.t. the hyperparameter $k$;
\item Application of the conformalization can significantly robustify and improve an anomaly detection method performance, cf. performance of LOF with that of LOF-ICAD;
\item KNN-CAD, although it does not use any predictive time-series model, is close in terms of performance to Numenta HTM, which is based on a predictive model. Therefore, there is a significant room for further performance improvement of the proposed method.
\end{enumerate}

\begin{table}[t!]
\caption{NAB Scoreboard}
\begin{center}
\onehalfspacing
\footnotesize{
\centering
\begin{tabular}{|l|c|c|c|}
\hline
\multirow{2}*{Detector} & \multicolumn{3}{c|}{Scores for Application Profiles} \\

& \textbf{Standard} & \textbf{Low FP} & \textbf{Low FN} \\
\hline
Numenta HTM & 65.3 & 58.6 & 69.4 \\
{\bf KNN-ICAD} & {\bf 57.99} & {\bf 43.41} & {\bf 64.81}  \\
Twitter ADVec & 47.1 & 33.6 & 53.5 \\
{\bf LOF-ICAD} & {\bf 36.7} & {\bf 30.12} & {\bf 42.11} \\
LoOP    & 14.63 & 8.47 & 24.7  \\
LOF    & 6.39 & 1.57 & 9.82  \\
Null & 0.0 & 0.0 & 0.0 \\
\hline
\end{tabular}
}
\end{center}
\label{tab:results}
\end{table}
\linespread{1.1}

\section{Conclusions}
We proposed non-parametric anomaly detection methods, suited both for a one-dimensional time-series data and a multi-dimensional data. Results of experiments provide evidence of high-competitiveness and beneficial properties of our methods.
\\\indent Further, we are going to extend the proposed methods to incorporate a time-series predictive model and to take into account properties of a manifold \cite{c1,c2}, approximating feature vectors \eqref{eq:ssa}.

\section*{Acknowledgements}
The research, presented in Section \ref{experiments} of this paper, was supported by the RFBR grants 16-01-00576 A and 16-29-09649 ofi\_m; the research, presented in other sections, was conducted in IITP RAS and supported solely by the Russian Science Foundation grant (project 14-50-00150).

\bibliographystyle{alpha}
\bibliography{bib_links}

\end{document}